\renewcommand{\thefootnote}{\fnsymbol{footnote}}

\bigskip
\documentclass[final,fleqn,times,twocolumn]{elsarticle}
\usepackage{graphicx}
\usepackage{fancyhdr}
\usepackage{dcolumn}
\usepackage{amssymb}
\usepackage{amsmath}
\usepackage[normalem]{ulem}
\usepackage{bm}
\usepackage{textcomp}
  \usepackage{natbib}
  \usepackage[a4paper,left=1.55cm,top=2cm,right=1.55cm]{geometry}
\usepackage[colorlinks=true,linkcolor=blue,citecolor=blue]{hyperref}
\usepackage[dvips]{color}
\newcommand{\com}[1]{{\sf\color[rgb]{0,0,1}{#1}}}

\journal{Physics Letters B}

\begin{document}

\begin{frontmatter}

\title{Correlation of the hyperon potential stiffness with hyperon constituents in neutron stars and heavy-ion collisions}
\author{Si-Na Wei$^{1,}$\footnotemark[1], Zhao-Qing Feng$^{1,}$\footnotemark[2], Wei-Zhou Jiang$^{2,}$\footnotemark[3]}
\address{$^1$School of Physics and Optoelectronics, South China University of Technology, Guangzhou
510640, China\\
$^2$School of Physics, Southeast University, Nanjing 211189, China}
\begin{abstract}
The breaking of SU(6) symmetry to a more general flavor SU(3) symmetry could serve as a potential explanation for the "hyperon puzzle" of neutron stars by adjusting the hyperon potentials. Specifically, when the soft relativistic mean-field (RMF) $\Lambda$ hyperon potentials  fall within the domains of chiral SU(3) interactions NLO13 with two-body forces, the maximum mass of neutron stars is expected to be lower than 2.0 $M_\odot$, whereas it can exceed $2.0M_\odot$ if the RMF $\Lambda$ hyperon potentials are sufficiently stiff to be consistent with those  from chiral SU(3) interactions NLO13 with three-body forces. In our investigation involving these two types of $\Lambda$ hyperon potentials, we explore how the hyperon yields and flows are affected  in heavy-ion collisions. We find that the inclusion of hyperon potentials results in better agreement   of the $\Lambda$ directed flows with data but without clear differentiation in the stiffness of the hyperon potentials. Similarly negligent is  the disparity in the rapidity distributions of the $\Lambda$ collective flows predicted by the stiff and soft hyperon potentials. In contrast, the $\Lambda$ collective flows beyond the central rapidity region turn out to be sensitive to the stiffness of the RMF equation of state (EOS) with the preference of a  soft RMF EOS to a stiff one. Notably, the transverse momentum distributions of $\Lambda$ hyperon production are sensitive to both the stiffness of the RMF EOS and $\Lambda$ hyperon potential at high transverse momenta.
\end{abstract}


\end{frontmatter}

\renewcommand{\thefootnote}{\fnsymbol{footnote}}
\footnotetext[1]{Electronic address: 471272396@qq.com}
\footnotetext[2]{Corresponding author: fengzhq@scut.edu.cn}
\footnotetext[3]{Electronic address: wzjiang@seu.edu.cn}
\renewcommand{\thefootnote}{\arabic{footnote}}

\section{INTRODUCTION}
Neutron stars (NSs), the most compact visible objects in the universe, play a crucial role in understanding the properties of dense hadronic matter, such as the equation of state (EOS) and phase transitions, at densities far beyond nuclear saturation density. The Shapiro delay measurements have successfully pushed forward the discovery of the large-mass NSs. The radio pulsars J1614-2230, J0348+0432 and J0740+6620  were typically measured  to have the masses $1.908\pm0.016M_\odot$~\cite{hy1,hy2,hy3}, $2.01\pm0.04M_\odot$~\cite{hy4} and $2.08\pm0.07M_\odot$~\cite{hy5,hy6}, respectively. The EOS should be stiff enough to support a NS mass over 2.0 $M_\odot$ that usually disfavors the appearance of new degrees of freedom at the cost of the EOS softening. In deed, in the core of a NS around 2.0 $M_\odot$,
the high hadronic density  far beyond nuclear saturation density enables the emergence of hyperons according to the chemical equilibrium. The appearance of hyperons will soften the EOS~\cite{hy7,hy8}, resulting in significant reduction of the NS maximum mass to well below 2.0 $M_\odot$~\cite{hy1,sch06}. This is known as the "hyperon puzzle".

Many schemes with additional repulsive interactions can stiffen the EOS and reconcile the "hyperon puzzle". Additional repulsive interactions, including hyperon-nucleon (hyperon) two-body and three-body forces, is able to produce the NS mass over $2.0 M_\odot$~\cite{hy9,hy10,hy11,hy12,hy13,hy14,hy15,hy16,hy17,hy18,hy19,ahy1}.  The appearance of other degrees of freedom, such as the $\Delta$ isobar, meson condensates and the quark matter below the hyperon thresholds can invoke new mechanisms to keep the large NS maximum mass in the presence of the hyperons~\cite{hy20,hy21,hy22,hy23}.
Moreover, the appropriate density-dependence of the hyperonic interactions in  the relativistic mean field (RMF) approach can also stiffen the the EOS  at high density without violation of the $2M_\odot$ constraint~\cite{hy24,hy25,hy26}.
In this work, we consider in the RMF model with  additional repulsion of hyperon-nucleon (hyperon) two-body forces achieved by reformulating the vector meson exchange. The  additional hyperon-vector meson coupling  arises from relaxing the SU(6) symmetry to the flavor SU(3) symmetry~\cite{hy9}.   In the  flavor SU(3) symmetry, the  hyperon-vector meson coupling strength can be rebuilt strong enough to raise the hyperon threshold  and reduce the hyperon fraction in NS,  supporting a large NS mass over $2.0 M_\odot$. In this flavor SU(3) model, the hyperon-scalar meson coupling is  determined through fitting the empirical hyperon potential depth at saturation. For instance, the depth of the $\Lambda$ hyperon potential $U_\Lambda$ is approximately 30 MeV at saturation density. Albeit primarily constrained by the $2.0 M_\odot$ NS mass,  the  hyperon potentials at high density still have significant uncertainty due to the lack of the direct constraint.

The energetic heavy-ion collisions are the unique way to extract the direct constraint on the in-medium hyperon potentials. The transverse and rapidity distributions of hyperon production and  collective flows  associated with the hyperon potential  have been widely studied both theoretically and experimentally in heavy-ion collisions~\cite{hy27,hy28,hy29,hy30,hy31,hy32,hy33,hy34,hy35,hy36,hy66,hy67,feng23}.
The quantum molecular dynamics (QMD) can successfully simulate the heavy-ion collisions. Since the threshold of hyperon production  is around the magnitude of \com{1.5 GeV}  in heavy-ion collisions, considering the relativistic effects becomes necessary. In the previous relativistic QMD models, the $\Lambda$ hyperon potential remains non-relativistic~\cite{hy36}. To address this issue, we  incorporate the RMF potentials into the relativistic QMD.  A more general approach is to incorporate the $\Lambda$  hyperon potential of  flavor SU(3) model into relativistic QMD. By doing so, "hyperon puzzle" in NS can be naturally associated with the properties of hyperons in heavy-ion collisions. Analyzing the collective flows and yields of the hyperon constituents in heavy-ion collisions may provide direct constraint on the stiffness of high-density hyperon potentials.

\section{Formalism}
\subsection{ RMF approach for neutron stars}
In RMF models, the interaction between baryons is achieved by exchanging mesons, such as the isoscalar-scalar $\sigma$, isovector-vector $\rho$,   and isoscalar-vector $\omega$ mesons.  The isoscalar-scalar $\sigma$  meson provides the medium-range attraction between baryons, while the isoscalar-vector $\omega$ and isovector-vector $\rho$ and $\phi$ mesons provide short-range repulsion. The interacting Lagrangian is given as~\cite{hy8,hy9,feng24,hy37}
\begin{eqnarray}
&\mathcal{L}_{int}=&\sum_B\bar{\psi}_{B}[g_{{B}\sigma}\sigma-\gamma_{\mu}(g_{{B}\omega}{\omega}^{\mu}+g_{{B}\phi}{\phi}^{\mu}+g_{{B}\rho}\vec{\tau}\cdot \vec{b}^{\mu})\nonumber\\
&& ]\psi_B-\frac{1}{3}g_2\sigma^3-\frac{1}{4}g_3\sigma^4,
\label{pos1}
\end{eqnarray}
where the sum on $B$ runs over  all the baryon octet ($p$, $n$, $\Lambda$, $\Sigma^+$, $\Sigma^0$, $\Sigma^-$, $\Xi^0$, $\Xi^-$),  the coupling constants for $\sigma, \omega, \rho,$ and $\phi$ mesons are denoted as $g_{{B}i}$ ($i=\sigma, \omega, \rho, \phi$), and  $g_2$ and $g_3$ are the coupling constants of the nonlinear self-interaction of isoscalar-scalar $\sigma$ meson.

For nuclear matter, we  adopt  the GM1 parametrizations~\cite{hy37}. While the hadron masses are the same as those in Ref.~\cite{hy37}, the meson coupling constants with nucleons and the self-interacting coupling constants of the $\sigma$ mesons are tabulated in Table~\ref{tb1}. In GM1,  the saturation properties are given as saturation density $\rho_0=0.153 fm^{-3}$,  binding energy $E/A(\rho_0)-M_N=-16.3$ MeV,  and incompressibility $K=300$ MeV, along with the symmetry energy $E_{sym}(\rho_0)$=32.5 MeV and slope of symmetry energy $L(\rho_0)$=94.0 MeV at saturation density. Due to the relatively high incompressibility of GM1, we also fit  two new sets of this work (\com{SOT1 and SOT2}). \com{The incompressibilities of SOT1 and SOT2 are $K=260$ MeV and $K=240$ MeV, respectively.}
The hadron masses, saturation density and  binding energy in \com{SOT1 and SOT2} are the same as those in GM1. \com{The  symmetry energies of both SOT1 and SOT2 are set to be 31.6 MeV~\cite{hy38}, with their slopes at saturation density being  87.8 MeV and 87.1 MeV, respectively.} The remaining parameters of  \com{SOT1 and SOT2} are  given in Table~\ref{tb1}.

\com{Due to the small number of hyperons produced in each collision event of QMD model, this study disregards the strangeness meson $\sigma^*$ which only mediates hyperon-hyperon interactions.  However, due to the breaking of SU(6) symmetry, the strangeness meson $\phi$  will interact with nucleons. Therefore, we preserve the strangeness meson $\phi$ . } When the $\phi$ meson is present, the model is named as $\sigma\omega\rho\phi$ models~\cite{hy9,hy39,hy40}. The mass of $\phi$ meson of this work is taken to be 1020 MeV~\cite{ahy2}. The breaking of SU(6) symmetry to the flavor SU(3) symmetry relaxes some constraints on the parameters, and three free parameters can appear in the flavor SU(3) symmetry:  the weight factor for the contributions of the symmetric and the antisymmetric  couplings $\alpha$,  the ratio between the meson octet   and singlet couplings $z$, and the mixing angle $\theta_V $ of mixing the meson octet and singlet. Since  the mixing angle obtained from the quadratic mass formula
for mesons $\theta_V \approx 40^\circ$ is close to the ideal mixing value ~\cite{hy41,hy42}, we  take the ideal mixing value $\theta_V$=tan$^{-1}(1/\sqrt{2})$ as in previous studies~\cite{hy9,hy12,hy13}. When one of  the remaining parameters ($z$ and $\alpha$) is fixed to its  SU(6) value, the other can be treated as a free parameter. By varying the free parameter, the coupling strengthes of nucleon-$\phi$ and hyperon-vector meson can be adjusted. In this work, we take the weight factor $\alpha$ as its SU(6) value ($\alpha=1$) and vary the ratio $z$. The coupling constants of nucleon-$\phi$ and hyperon-vector meson are then obtained as~\cite{hy9}
\begin{eqnarray}
&&\frac{g_{\Lambda\omega}}{g_{N\omega}}=\frac{g_{\Sigma\omega}}{g_{N\omega}}
=\frac{\sqrt{2}}{\sqrt{2}+\sqrt{3}z},\nonumber\\&&
\frac{g_{\Lambda\phi}}{g_{N\omega}}=\frac{g_{\Sigma\phi}}{g_{N\omega}}
=\frac{-1}{\sqrt{2}+\sqrt{3}z},\nonumber\\&&
\frac{g_{\Xi\omega}}{g_{N\omega}}=
\frac{\sqrt{2}-\sqrt{3}z}{\sqrt{2}+\sqrt{3}z},\nonumber\\&&
\frac{g_{\Xi\phi}}{g_{N\omega}}=
-\frac{1+\sqrt{6}z}{\sqrt{2}+\sqrt{3}z},\nonumber\\&&
\frac{g_{N\phi}}{g_{N\omega}}=
-\frac{\sqrt{6}z-1}{\sqrt{2}+\sqrt{3}z}.
\label{pos2}
\end{eqnarray}

When the ratio $z$ takes its SU(6) value $z=1/\sqrt{6}$, the $\phi$, which is  a pure $\bar{s}s$ state,  does not couple to the nucleon. For $z\neq1/\sqrt{6}$, the $\phi$ will couple to the nucleon, and  contribute to the saturation properties of nuclear matter.   \com{To make sure that the same saturation properties of nuclear matter with and without $\phi$ mesons, one needs to make the following replacement~\cite{hy9}:}
\begin{eqnarray}
\frac{\tilde{g}^2_{N\omega}}{m_\omega^2}&& \Rightarrow
\frac{{g}^2_{N\omega}}{m_\omega^2}+\frac{{g}^2_{N\phi}}{m_\phi^2}\nonumber\\&&
=\frac{{g}^2_{N\omega}}{m_\omega^2}+
(\frac{\sqrt{6}z-1}{\sqrt{2}+\sqrt{3}z})^2\frac{{g}^2_{N\omega}}{m_\phi^2},
\label{pos3}
\end{eqnarray}
\com{where $\tilde{g}_{N\omega}$ is the value of GM1, SOT1 or SOT2 models without the $\phi$ meson. When the ratio $z$ ($\neq1/\sqrt{6}$) is given, since the $\phi$ will couple to the nucleon, the ${g}_{N\omega}$ is determined in such a way that the right-hand side of Eq.(\ref{pos3}) reproduces the same saturation properties of the \com{GM1, SOT1 or SOT2} models without the $\phi$ meson, and shown in Table~\ref{tb1}.}
After obtaining ${g}_{N\omega}$, the coupling constants of nucleon-$\phi$ and hyperon-vector meson are given through Eq.(\ref{pos2}). The  coupling constants of hyperon-scalar meson are  obtained through fitting the empirical hyperon potential depth. The hyperon potential is defined as~\cite{hy43,hy44}
\begin{eqnarray}
U_Y(\rho_B)=g_{Y\omega}\omega_0(\rho_B)+g_{Y\phi}\phi_0(\rho_B)-g_{Y\sigma}\sigma_0(\rho_B),
\label{pos4}
\end{eqnarray}
where the $\omega_0$, $\phi_0$ and $\sigma_0$ are the meson fields in RMF approximation, and $\rho_B$ is the baryon density. The hyperon potential depths $U_Y(\rho_0)$  are chosen as the empirical ones~\cite{hy45,hy46,hy47,hy48,hy49,hy50,hy51}:
\begin{eqnarray}
U_\Lambda(\rho_0)=-U_\Sigma(\rho_0)=-30 \rm{MeV}, U_\Xi(\rho_0)=-14  \rm{MeV}.
\label{pos5}
\end{eqnarray}

\begin{table*}[!htp]
\begin{center}
\caption{  Parametrizations  of GM1 and the set of this work (\com{SOT1 and SOT2}). Hadron masses are the same as those in Ref.~\cite{hy37}.  \label{tb1}}
\footnotesize
\begin{tabular*}{170mm}{c|@{\extracolsep{\fill}}c|c|c|c|c|c|c|c|c}
\hline
  model&   $g_{N\sigma}$  &    $\tilde{g}_{N\omega}$ &   ${g}_{N\omega}$  &   $g_2$ ($fm^{-1}$) &   $g_3$ &   $g_{N\rho}$  &  $K$ (MeV) & $E_{sym}$($\rho_0$) (MeV) & $L$ ($\rho_0$)(MeV) \\
\hline
GM1(z=0.4)  & 8.8950 & 10.6100 &10.6097  &  9.8589  & -6.6984&4.0975& 300& 32.5 &94.0 \\
GM1(z=0.8)  & 8.8950 & 10.6100 &10.2608  &  9.8589  & -6.6984&4.0975& 300& 32.5 &94.0 \\
SOT1(z=0.1)  & 8.3000 & 9.2030  & 8.6448  &16.7500  & -4.0700 &4.1240& 260&31.6&87.8  \\
SOT1(z=0.6)  & 8.3000 & 9.2030  & 9.1052  &16.7500  & -4.0700 &4.1240& 260&31.6&87.8  \\
SOT2(z=0)  & 8.1760 & 8.7460  & 7.6866  &22.2000  & -11.5000 &4.1610& 240&31.6&87.1  \\
SOT2(z=0.5)  & 8.1760 & 8.7460  & 8.7211  &22.2000  & -11.5000 &4.1610& 240&31.6&87.1  \\
\hline
\end{tabular*}
\end{center}
\end{table*}

When the baryon-meson coupling constants  are fixed, the meson fields in asymmetric baryonic matter in the RMF approximation can be obtained from solving the Euler-Lagrange equations
with the imposition of the chemical equilibrium and charge neutral condition~\cite{zh11,xi14,wei21,wei22}.  Given the  meson fields,  one can obtain the energy density and pressure in asymmetric baryonic matter that are necessary inputs of the Tolman-Oppenheimer-Volkoff (TOV) equations~\cite{hy52,hy53}. The mass-radius trajectories of NSs are then obtained by solving the TOV equations.

\subsection{ RMF approach in heavy-ion collisions}
In a relativistic N-body system, both the position coordinates $q_{i}^{\mu}$ and the momentum coordinates $p_{i}^{\mu}$ have  4$N$ dimensions.
The on-mass shell conditions and  the time fixation constraints are
needed to reduce the number of dimensions from 8$N$ to physical trajectories 6$N$~\cite{hy54,hy55}.  The Hamiltonian of the $N$-body system can be constructed through  the linear combination of $2N-1$ constraints~\cite{hy36,hy56,hy57,hy58,hy59,hy60}:
\begin{eqnarray}
H=\sum_{i=1}^{2N-1}\lambda_i(\tau)\Phi_i,
\label{rqmd1}
\end{eqnarray}
where $\Phi_i$ is on-mass shell conditions $\Phi_i\equiv p_i^{*2}-M_i^{*2}=(p_i-V_i)^2-(M_B-S_i)^2$. Assuming  $[\Phi_i,\Phi_j]=0$,  it follows  that the $\lambda_i=0$ for $N+1<i<2N$.  When the time fixation constraints  obey the world-line condition, the $\lambda_i$ is obtained as $\lambda_i=1/(2p_i^{*0})$~\cite{hy59,hy60}.
The evolution equations of the coordinates can be obtained from   the Hamiltonian as~\cite{hy36,hy57,hy58}
\begin{eqnarray}
&&\dot{\vec{r}}_i=\frac{\vec{p}_i^*}{p_i^{*0}}+\sum_{j=1}^N(\frac{M_j^*}{p_j^{*0}}\frac{\partial M_j^*}{\partial \vec{p_i}}+z_j^{*\mu}\cdot\frac{\partial V_{j\mu}}{\partial\vec{p_i}}),\nonumber\\&&
\dot{\vec{p}}_i=-\sum_{j=1}^N(\frac{M_j^*}{p_j^{*0}}\frac{\partial M_j^*}{\partial \vec{r_i}}+z_j^{*\mu}\cdot\frac{\partial V_{j\mu}}{\partial\vec{r_i}}),
\label{rqmd2}
\end{eqnarray}
where $z_i^{*\mu}=p_i^{*\mu}/p_i^{*0}$,  $p_{i,\mu}^*=p_{i,\mu}-V_{i,\mu}$, and $M_i^*=M_B-S_i$.
The scalar potential $S_i$ and vector potential $V_{i,\mu}$  are in turn  defined as
\begin{eqnarray}
&&S_i=g_{B\sigma}\sigma_i,\nonumber\\&&
V_{i,\mu}={B_i}g_{B\omega}\omega_{i,\mu}+{B_i}g_{B\phi}\phi_{i,\mu}+{B_it_i}g_{B\rho} b_{i,\mu},
\label{rqmd3}
\end{eqnarray}
where $B_i$ is the baryon number of the $i$th particle and
$t_i=\pm 1$ are the positive and negative projections of the third component of isospin, respectively.
The meson fields are just the  RMF ones associated with the dense matter in the collision:
\begin{eqnarray}
&&m_\sigma^2\sigma_i+g_2\sigma_i^2+g_3\sigma_i^3=g_{B\sigma}\rho_{S,i},\nonumber\\&&
m_\omega^2\omega_{i,\mu}=g_{B\omega} J_{i,\mu},\nonumber\\&&
m_\phi^2\phi_{i,\mu}=g_{B\phi} J_{i,\mu},\nonumber\\&&
m_\rho^2b_{i,\mu}=g_{B\rho} R_{i,\mu},
\label{rqmd4}
\end{eqnarray}
with
\begin{eqnarray}
&&\rho_{S,i}=\sum_{j\neq i}\frac{M_j}{p_j^0}\rho_{ij},\qquad
J_{i,\mu}=\sum_{j\neq i}B_j\frac{p_{j,\mu}}{p_j^0}\rho_{ij},\nonumber\\&& R_{i,\mu}=\sum_{j\neq i}t_jB_j\frac{p_{j,\mu}}{p_j^0}\rho_{ij}.
\label{rqmd5}
\end{eqnarray}
Here, the interaction density in  relativistic QMD is defined as $\rho_{ij}=\frac{\gamma_{ij}}{(4\pi L_G)^{3/2}}\mathrm{exp}(\frac{q_{T,ij}^2}{4L_G})$ with $L_G=2.0$ $\rm{fm^2}$ and  $q_{T,ij}^2= q_{ij}^2-[\frac{q_{ij,\sigma}(p_i^\sigma+p_j^\sigma)}{\sqrt{(p_i+p_j)^2}}]^2$. The Lorentz factor $\gamma_{ij}$, which  is expressed as $(p_i^0+p_j^0)/(p_i+p_j)$  in two-body center-of-mass frame, ensures the correct normalization of the Gaussian~\cite{rmq70}.

With the on-mass shell conditions, the hyperon potential in   relativistic QMD  can be obtained from the in-medium dispersion relation:
\begin{eqnarray}
U_Y=\sqrt{(M_Y+V_i)^2+\vec{p}_i^2}-S_i-M_Y,
\label{rqmd6}
\end{eqnarray}
where $M_Y$ is the free hyperon mass. When $\vec{p}$ takes zero, the hyperon potential in symmetric nuclear matter  can be simplified as
\begin{eqnarray}
U_Y=V_i-S_i={B_i}g_{B\omega}\omega_{i,\mu}+{B_i}g_{B\phi}\phi_{i,\mu}-g_{B\sigma}\sigma_i,
\label{rqmd7}
\end{eqnarray}
which is the same as Eq.(\ref{pos4}). This suggests that, in addition to the NS properties, the heavy-ion collisions may be used to discriminate the stiffness of high-density hyperon potential.

\section{Results and discussions}
\subsection{Hypernuclear matter and neutron stars}

\begin{figure}
\centerline{\includegraphics[height=6cm,width=8cm]{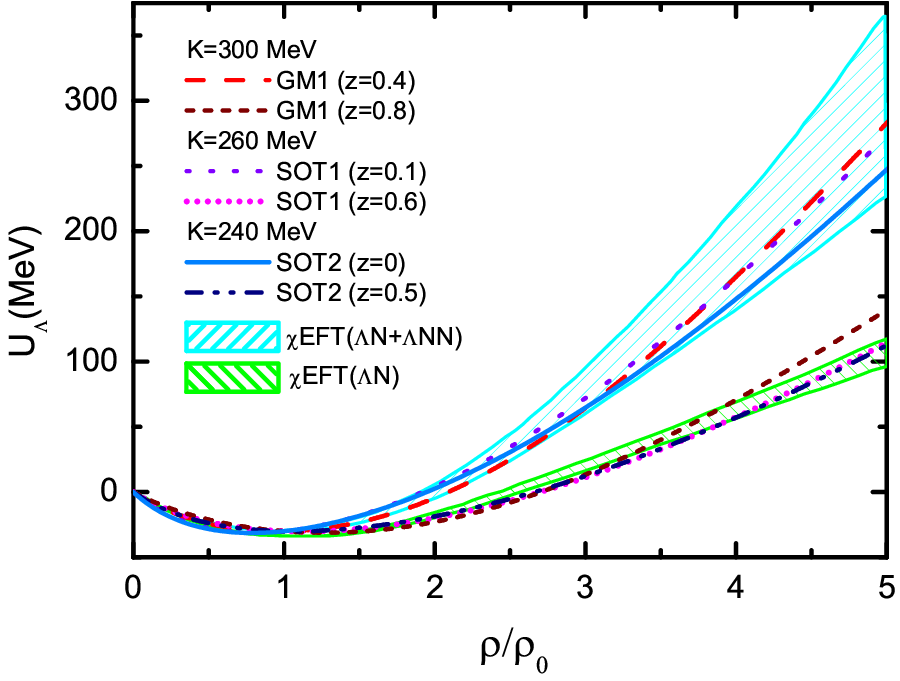}}
\caption{(Color online)   The $\Lambda$ hyperon potentials $U_\Lambda$ of symmetric nuclear matter  in $\sigma\omega\rho\phi$  model with  various values of $z$. The cyan and green areas are the results of the SU(3) chiral effective field theory ($\chi{EFT}$) with two- ($\Lambda N$) and three-body ($\Lambda N+\Lambda NN$) forces~\cite{ahyd}, respectively.}\label{flam}
\end{figure}

\begin{figure}[h]
\centerline{\includegraphics[height=6cm,width=8cm]{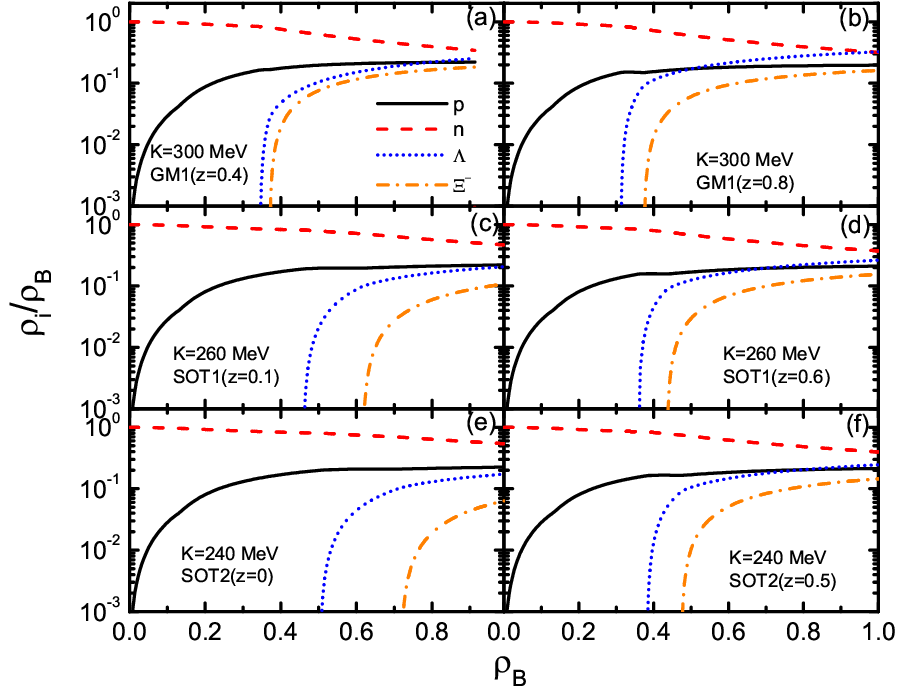}}
\caption{(Color online)  Baryon fractions as a function of total baryonic density in the $\sigma\omega\rho\phi$  model with various values of $z$. }\label{fracl}
\end{figure}
To manifest the effect of the stiffness of the in-medium hyperon potentials, we firstly construct two  hyperon potentials with different stiffness by varying the parameter $z$ which  is the ratio of the meson octet coupling to that of the singlet in the flavor SU(3) symmetry.
As shown in Fig.~\ref{flam}, the stiff hyperon potentials for symmetric nuclear matter with \com{$z=0.4$ in GM1, $z=0.1$ in SOT1 and $z=0$ in SOT2,} obtained from Eq.(\ref{pos4}),  align with the results of the SU(3) chiral effective field theory ($\chi{EFT}$)  with three-body forces, while the soft hyperon potentials with \com{$z=0.8$ in GM1, $z=0.6$ in SOT1 and $z=0.5$ in SOT2}  well simulates the results of $\chi{EFT}$  with two-body forces. With or without these hyperon potentials, we can simulate various cases to capture the signals sensitive to the hypernuclear EOS in NSs.

\begin{figure}
\centerline{\includegraphics[height=6cm,width=8cm]{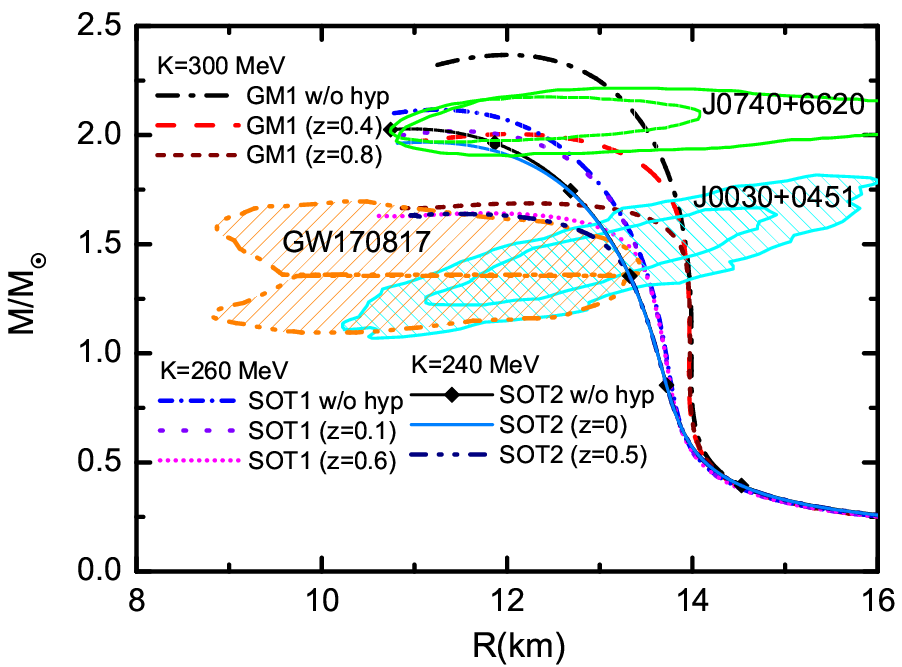}}
\caption{(Color online)  The mass-radius trajectories of NSs for the matter of nucleon doublet and  baryon
octet in $\sigma\omega\rho\phi$  model. The  regions with orange dash dot dot line are the constraints extracted from the GW170817 event~\cite{hy61,hy62}. \com{The regions with cyan solid line and green solid line are the constraints  of the NICER  observations for   J0030+0451~\cite{hy63,hy64} and J0740+6620~\cite{hy65,hym66}, respectively.}  }\label{fns}
\end{figure}

The hyperon \com{in NS} can arise at various densities exceeding 0.3 $fm^{-3}$, as shown in Fig.~\ref{fracl}, depending on the specific value of the ratio parameter $z$.  Notably,  Fig.~\ref{fracl} also shows that the predominant hyperonic component is the $\Lambda$ hyperon, while the $\Sigma$ hyperons, with a positive hyperon potential depth, do not appear. \com{For a given hyperon potential (within the domain of three-body forces or two-body forces), the softer the EOS of RMF, the higher the baryon density of hyperons appearing.}
Within a specific model \com{(GM1, SOT1 or SOT2)}, both the onset density of the hyperons and  the stiffness  of  hypernuclear EOS rely notably on the parameter $z$.  It is apparent that a larger value of $z$ will lead to a lower onset density for hyperons within a given model. Consequently, a larger value of  $z$ results in a softer EOS of hypernuclear matter in NS.
With the EOSs of pure nuclear matter and hypernuclear matter,  the radius-mass trajectories of static NSs  can be derived by solving the standard TOV equation.
\com{Since the crust-core transition density of NS using RMF model is approximately  0.08 $fm^{-3}$~\cite{crust1,crust2}, we have employed the  EoS of RMF from this work  for densities above 0.08 $fm^{-3}$ in TOV equation~\cite{hy26}. For densities below 0.08 $fm^{-3}$,  we  have adopted the EOS of Ref.\cite{bay71}  for densities less than $2.57\times10^{-4}$ $fm^{-3}$  and the EOS of Ref.\cite{iid76} for densities between $2.57\times10^{-4}$ $fm^{-3}$  and $0.08$ $fm^{-3}$.
As shown in Fig.~\ref{fns}, the NS mass-radius trajectories using the GM1, SOT1, and SOT2 models all align with the radius constraints derived from the NICER observations for   J0030+0451~\cite{hy63,hy64}, and they are positioned within or very close to the marginal regions extracted from  GW170817 event~\cite{hy61,hy62}.}
The maximum mass of NSs with pure nuclear matter EOS are $2.37 M_\odot$, $2.12 M_\odot$ and $2.03 M_\odot$ for  the \com{GM1, SOT1 and SOT2,} respectively,  both exceeding $2.0M_\odot$. In the presence of hyperons in NS, the NS maximum mass  is considerably reduced. Since the ratio parameter $z$ dictates  the stiffness of hypernuclear EOS, the NS maximum mass can increase significantly with the  decrease of $z$.
\com{The NS maximum mass with $z=0.4$ in GM1, $z=0.1$ in SOT1 and $z=0$ in SOT2  are $2.00 M_\odot$, $2.02 M_\odot$ and $1.97 M_\odot$, respectively, aligning with  the results  of the NICER  observations for J0740+6620~\cite{hy65,hym66}. However, the NS maximum mass with  $z=0.8$ in GM1, $z=0.6$ in SOT1 and $z=0.5$ in SOT2 are  notably  lower than $2.0 M_\odot$  and do not meet the findings  of the NICER  observations for  J0740+6620.}
Apart from the NS maximum mass, we expect to offer insights into the in-medium hyperon potentials directly from analyzing heavy-ion collisions.

\subsection{Collective flows and yields of $\Lambda$ hyperon}

The predominant hyperonic component is the $\Lambda$ hyperon either in asymmetric nuclear matter or in the production by the collisions for the smallest threshold. At saturation density, the $\Lambda$ hyperon potential $U_\Lambda$, approximately around -30 MeV, is well constrained by the properties of single and double $\Lambda$ hypernuclei. However, due to the limited constraints on the in-medium baryon-baryon interactions associated with abundant intermediates, considerable uncertainty may exist in the $\Lambda$ hyperon potential at high density. Apart from the NS maximum mass, it is necessary to check the sensitivity of observables to the hyperon potentials directly from the heavy-ion collisions in which the  strange particles, including hyperons, are predominantly  produced through the  inelastic hadron-hadron collisions. The channels of strangeness productions are taken as the same as the \com{Lanzhou quantum molecular dynamics} (LQMD) model~\cite{hy35,hy66,hy67,feng23}:
\begin{eqnarray}
&&BB\rightarrow{BYK},BB\rightarrow{BBK\overline{K}},B\pi\rightarrow{NK\overline{K}},\nonumber\\&&
{YK}\rightarrow{B\pi},B\pi(\eta)\rightarrow{YK},Y\pi\rightarrow{B\overline{K}}, B\overline{K}\rightarrow{Y\pi},\nonumber\\&&
YN\rightarrow{\overline{K}NN},BB\rightarrow{B\Xi KK},\overline{K}B\leftrightarrow{K\Xi},\nonumber\\&&
YY\leftrightarrow{N\Xi},\overline{K}Y\leftrightarrow{\pi\Xi},
\label{chan1}
\end{eqnarray}
where the symbol correspondences are given as B ($N, N^*,\Delta$),   $Y$ ($\Lambda, \Sigma$), $\Xi$ ($\Xi^0, \Xi^-$), $\pi$ ($\pi^-,\pi^0,\pi^+$), $K$ ($K^0,K^+$), and $\overline{K}$ ($\overline{K}^0,K^-$). Indicative of  the strength of interaction, the collective flows of  $\Lambda$  hyperons have been investigated experimentally~\cite{hy27,hy28,hy29,hy30,hy31,hy32,hy33,hy34}. Aimed at extracting the constraint on hyperon potentials, the collective flows of $\Lambda$ hyperons have been predicted with various hyperon potentials~\cite{hy35,hy36}. In this work, we aim to investigate how various $\Lambda$ hyperon potentials allowed by the flavor  SU(3) symmetry impact collective flows and yields of $\Lambda$  hyperons.

\begin{figure}
\centerline{\includegraphics[height=8cm,width=8cm]{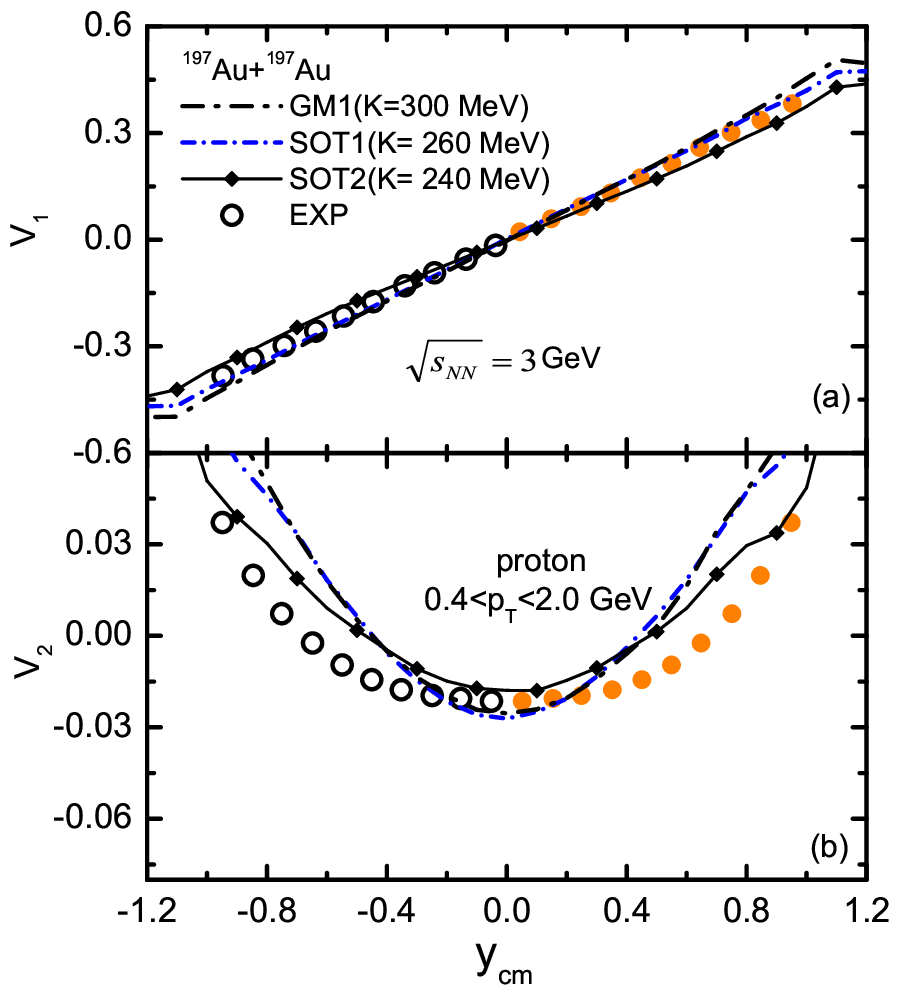}}
\caption{(Color online)   Rapidity distribution of the collective
flows of proton at the nucleon-nucleon center-of-mass energy $\sqrt{s_{NN}}=3$ GeV for the near-central ($b=1-3$ fm) $^{197}$Au+$^{197}$Au reactions. \com{The black open  circles represent the experimental data of the STAR
collaboration~\cite{hy33}, while the orange solid  circles represent the experimental data after symmetric translation.}}\label{fpro}
\end{figure}

Firstly, we investigate the directed and elliptic flows of protons  with  \com{GM1, SOT1 and SOT2}. As shown in upper panel of Fig.~\ref{fpro}, the  directed flows of protons with  \com{GM1, SOT1 and SOT2} agree well with  the experimental data of STAR collaboration  across the entire rapidity~\cite{hy33}. As compared to the fit  with the non-relativistic QMD with Skyrme interactions, the satisfactorily  improved agreement here can be ascribed primarily to the Lorentz effect in the relativistic QMD~\cite{hy54,wei23}.
As shown in lower panel of Fig.~\ref{fpro}, \com{the elliptic flows of  protons with  GM1,  SOT1 and SOT2 are only consistent with the experimental data only at small rapidity.
At large rapidity, the elliptic flows of protons predicted by the GM1 and SOT1 models increase more rapidly than the experimental data. However, the elliptic flows of protons from the SOT2 model closely align with the experimental data, showing a closer match compared to those predicted by the GM1 and SOT1 models. Additionally, at large rapidity, the curvature of the elliptic flows predicted by the SOT2 model closely resembles that of the experimental data.}
The deviation \com{between the elliptic flows of GM1 and SOT1 models and  the experimental data} is actually associated with the EOS stiffness at relatively lower densities corresponding to the large rapidity region~\cite{wei23}.

\begin{figure}
\centerline{\includegraphics[height=8cm,width=8cm]{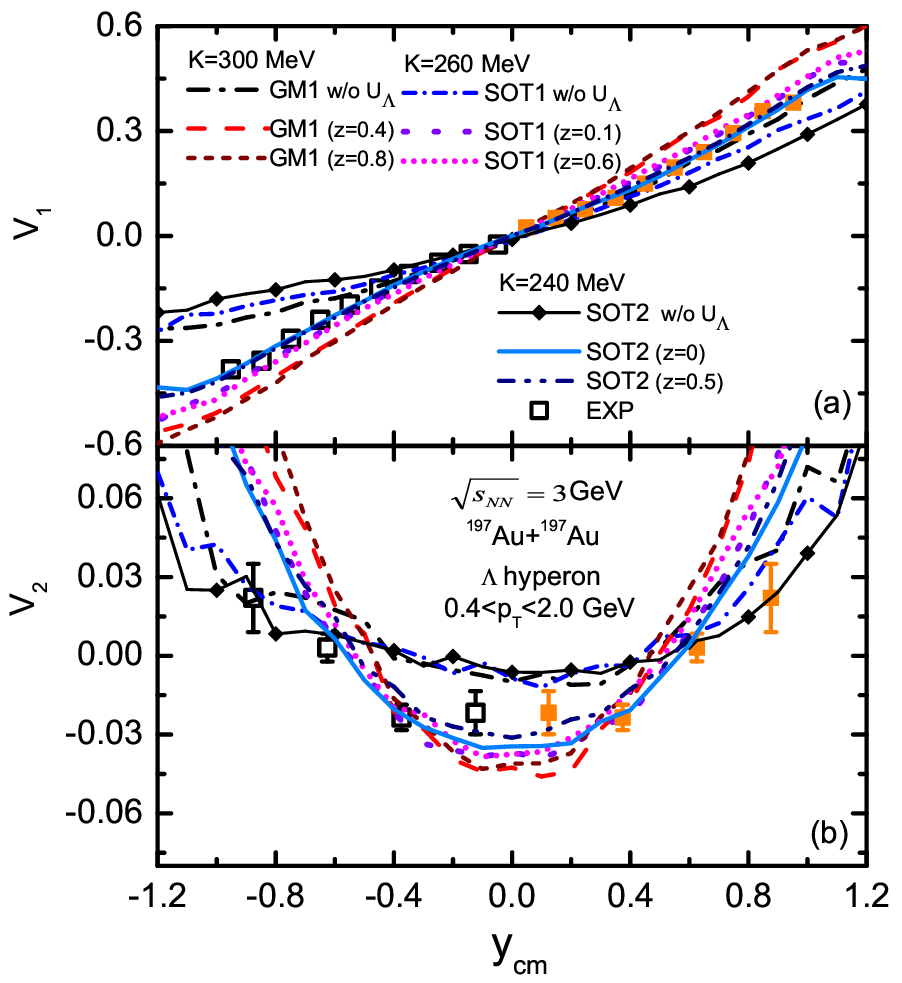}}
\caption{(Color online)  Rapidity distribution of the collective
flows of $\Lambda$ hyperon at the nucleon-nucleon center-of-mass energy $\sqrt{s_{NN}}=3$ GeV for the near-central ($b=1-3$ fm) $^{197}$Au+$^{197}$Au reactions. \com{The black open  squares represent the experimental data of the STAR collaboration~\cite{hy33}, while the orange solid  squares represent the experimental data after symmetric translation.}}\label{fhyp}
\end{figure}

The collective flows of protons reflect the interactions between nucleons, which similarly impact the production and motion of hyperons.
With the reasonable fit to the collective flows of protons in  the near-central ($b=1-3$ fm) $^{197}$Au+$^{197}$Au reactions, we proceed to study the collective flows of $\Lambda$ hyperon.
The directed and elliptic flows  of $\Lambda$ hyperon with various hyperon potentials for the near-central ($b=1-3$ fm) $^{197}$Au+$^{197}$Au reactions are shown in upper and lower panels of Fig.~\ref{fhyp}, respectively.
The $\Lambda$ directed flows with the GM1 and SOT1 without the hyperon potentials can well fit the experimental data from STAR collaboration~\cite{hy33} at  rapidity $|y_{cm}|\lesssim 0.6$. \com{However, the $\Lambda$ directed flows with  SOT2 without the hyperon potentials only match the experimental data at  rapidity $|y_{cm}|\lesssim 0.4$.
At large rapidity,  the $\Lambda$ directed flows from the GM1, SOT1 and SOT2  without the hyperon potentials  show poorer agreement with the experimental data.}
Besides, the $\Lambda$ directed flows with the GM1 without the hyperon potentials agree better with the data than those with the \com{SOT1 and SOT2} across the entire rapidity, which suggests that the difference in the incompressibility in these two models may account for  their deviation.
With the inclusion of the $\Lambda$ hyperon potentials,   the $\Lambda$ directed flows with the \com{SOT1 and SOT2} are totally  consistent with the  experimental data across the entire rapidity, while the deviation arises in the fit with the GM1.  In a given RMF model (\com{GM1, SOT1 or SOT2}), the $\Lambda$ directed flows with  the soft $\Lambda$  hyperon potentials are almost the same as those  with stiff $\Lambda$  hyperon potentials.
In other words,  the $\Lambda$ directed flows are not very sensitive to the density dependence of the $\Lambda$ hyperon potentials. This result is consistent with the findings of the Fermi momentum expansion of the potential~\cite{hy36}. An explanation of this result is that the average density is not high enough to  distinguish  the stiffness of the  $\Lambda$ hyperon potential, although the density is high in the center of heavy-ion collisions.
Since the  average density is sensitive to the nuclear EOS stiffness of RMF models,  the $\Lambda$ directed flows exhibit the sensitivity to the nuclear EOS stiffness, rather than the stiffness  of $\Lambda$ hyperon potentials. The $\Lambda$ directed flows prefer a soft EOS (\com{SOT1 or SOT2}) over a stiff EOS (GM1).
\com{Furthermore, it is important to emphasize that as protons exist in both low and high-density environments, the direct flow of protons is averaged across various densities, leading to a relatively weak correlation with the stiffness of the EOS. In contrast, hyperons are primarily produced at densities approximately 2.25 times the saturation density\cite{cheng22}. As a result, the direct flow of hyperons seems to demonstrate a more pronounced correlation with the stiffness of the EOS, compared to the direct flow of protons.}

\com{When hyperon potentials are included, we can observe from the lower panel of Fig.~\ref{fhyp} that the $\Lambda$ elliptic flows with a soft EOS  (SOT1 or SOT2) also tend to agree better with the experimental data than those with a stiff EOS (GM1). In particular, the $\Lambda$ elliptic flows of SOT2 closely match the experimental data at  rapidity $|y_{cm}|\lesssim 0.6$.  However, discrepancies between the $\Lambda$ elliptic flows and the experimental data exist for all models at large  rapidity. The discrepancies between the $\Lambda$ elliptic flows and the experimental data are still model-dependent~\cite{hy33}, possibly stemming from variations in the mechanisms of strangeness production across different models.}
\begin{figure}
\centerline{\includegraphics[height=8cm,width=8cm]{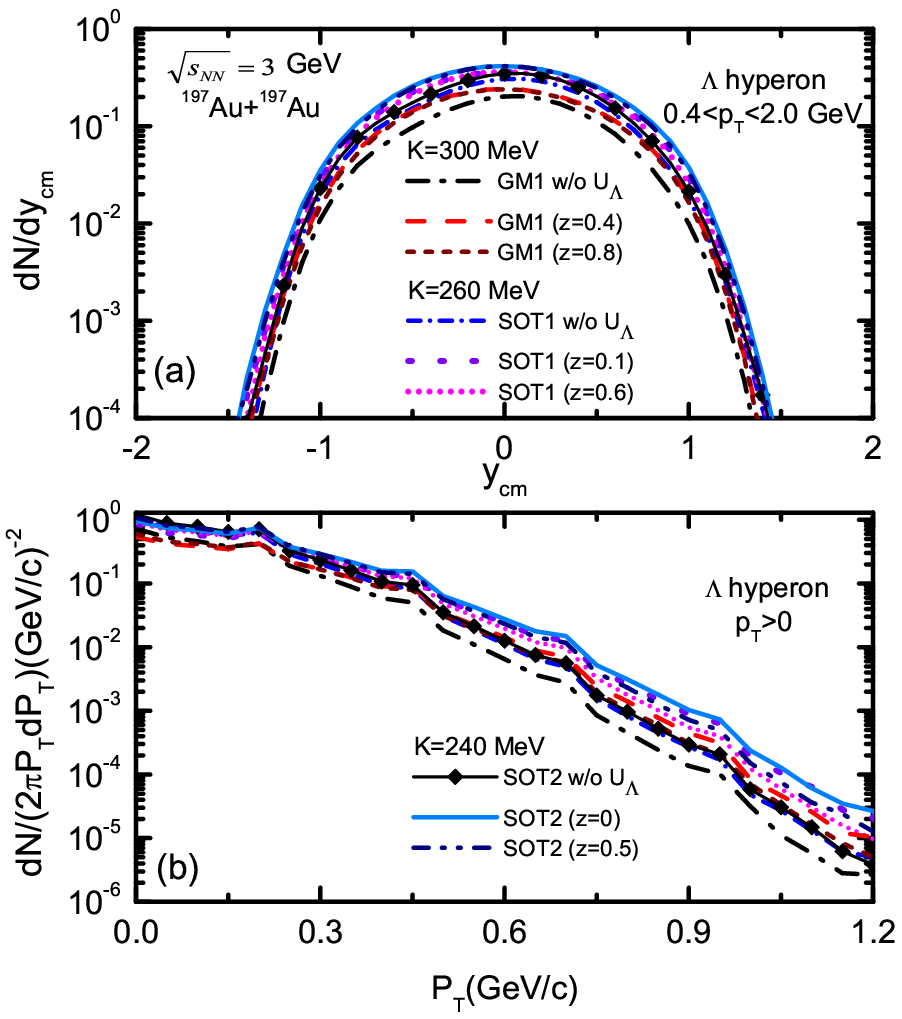}}
\caption{(Color online)  Rapidity (upper panel) and transverse momentum (lower panel) distributions of $\Lambda$  hyperon yields  at the nucleon-nucleon center-of-mass energy $\sqrt{s_{NN}}=3$ GeV  for the near-central ($b=1-3$ fm) $^{197}$Au+$^{197}$Au reactions. }\label{dnyp}
\end{figure}

On the other hand, the rapidity and transverse momentum distributions of $\Lambda$  hyperon yields are predicted to be influenced by the $\Lambda$  hyperon potentials. As depicted in the upper panel of Fig. \ref{dnyp}, the rapidity distributions of $\Lambda$  production  with $\Lambda$  hyperon potentials are higher than those without $\Lambda$  hyperon potentials. Similar to the collective flows, the rapidity distributions of $\Lambda$  production for a given RMF model (\com{GM1, SOT1 or SOT2}) are not very sensitive to the stiffness of the $\Lambda$  hyperon potentials but  sensitive to the stiffness of the EOS. In contrast, the transverse momentum distributions of the $\Lambda$  production  are sensitive not only to the stiffness of the EOS but also to the stiffness of the $\Lambda$  hyperon potential, as shown in the lower panel of Fig.~\ref{dnyp}. With similar $\Lambda$  hyperon potentials, a soft EOS  (\com{SOT1 or SOT2}) predicts a higher $\Lambda$ yield than the stiff EOS (GM1). As illustrated in the lower panel of Fig. \ref{dnyp}, for a given RMF model (\com{GM1, SOT1 or SOT2}), the transverse momentum distributions of $\Lambda$  yields  with the stiff $\Lambda$  hyperon potentials  are higher than those with soft $\Lambda$  hyperon potentials at large transverse momenta. The large transverse momentum constituents arise from the encounter of
the strong repulsions  at high densities.
The stronger repulsion with the stiff $\Lambda$  hyperon potentials at high densities thus makes it easier for $\Lambda$  hyperons to escape prior to being absorbed, resulting in the rise of their yields at large transverse momenta.

\section{Conclusions}

The relationship between the hyperon constituents in NS and  heavy-ion collision has been  studied with various ($\Lambda$) hyperon potentials  in conformity to the flavor SU(3) symmetry. The  presence of hyperon can  indeed reduces the NS maximum mass significantly. The stiff $\Lambda$ hyperon potentials, constructed  within those of the chiral SU(3) interactions NLO13  with three-body forces ($\Lambda N+\Lambda NN$), can  support the NS maximum mass above $2.0M_\odot$, while the soft $\Lambda$ hyperon potentials constructed within those of the chiral SU(3) interactions NLO13  with two-body forces ($\Lambda N$) can only lead to a NS maximum mass around $1.65M_\odot$.
Meanwhile, using both stiff and soft $\Lambda$  hyperon potentials, we investigate the directed and elliptic flows of protons and $\Lambda$  hyperons for near-central ($b=1-3$ fm) $^{197}$Au+$^{197}$Au reactions in comparison to the data of STAR collaboration. The relativistic and medium effects in terms of the nuclear EOS can well account for the proton collective flows. When employing the $\sigma\omega\rho\phi$ model (\com{GM1, SOT1 or SOT2}), it is observed that the directed flows of $\Lambda$ hyperons are notably sensitive to the stiffness of the RMF model whilst with minimal sensitivity to the stiffness of the hyperon potentials.  Specifically, the fit to the directed flows of $\Lambda$ hyperons prefers to softer model (\com{SOT1 or SOT2}) to the stiffer model (GM1). For the elliptic flows, it is similar to show a preference for the softer model.   Similar to the collective flows, the rapidity distributions of $\Lambda$ hyperon production are of clear sensitivity to the stiffness of the RMF EOS but with little sensitivity to the stiffness of the hyperon potentials. In contrast, the transverse momentum distributions of $\Lambda$ hyperon yields are sensitive not only to the stiffness of the RMF EOS, but also to the stiffness of the $\Lambda$ hyperon potential. Notably, for a given RMF model (\com{GM1, SOT1 or SOT2}), the transverse momentum distributions of $\Lambda$  yields  with the stiff $\Lambda$ hyperon potentials are higher than those with the soft $\Lambda$ hyperon potentials  at large transverse momenta. This provides a significant signal, yet to be verified experimentally, to be associated with the NS maximum mass observation in the presence of hyperon constituents.

\section*{ACKNOWLEDGMENTS}
This work was supported by the National Natural Science Foundation of China (Projects Nos. 12147106, 12175072, and 12375112) and the Talent Program of South China University of Technology (Projects No. 20210115).

\end{document}